\documentclass[prd,aps,preprint,nofootinbib,amssymb,eqsecnum]{revtex4}
\usepackage{amsmath}
\usepackage{verbatim,graphics,graphicx,xcolor,slashed}
\usepackage{ulem} 
\usepackage{hyperref}
\usepackage{changepage}

\newcommand{\tgray}[1]{\textcolor{gray}{#1}}

\newcommand{\ec}{\tgray{\rule[2pt]{2.2em}{2pt}}} 
\newcommand{\mcA}{\mathcal A}
\newcommand{\mcB}{\mathcal B}
\newcommand{\mcC}{\mathcal C}
\newcommand{\mcS}{\mathcal S}
\newcommand{\CP}{\text{CP}}

\def\lsim{\mathrel {\vcenter {\baselineskip 0pt \kern 0pt
    \hbox{$<$} \kern 0pt \hbox{$\sim$} }}}
\def\gsim{\mathrel {\vcenter {\baselineskip 0pt \kern 0pt
    \hbox{$>$} \kern 0pt \hbox{$\sim$} }}}
\def\slashchar#1{\setbox0=\hbox{$#1$}           
 \dimen0=\wd0                                 
  \setbox1=\hbox{/} \dimen1=\wd1               
\ifdim\dimen0>\dimen1                        
  \rlap{\hbox to \dimen0{\hfil/\hfil}}      
  #1                                        
  \else                                        
 \rlap{\hbox to \dimen1{\hfil$#1$\hfil}}   
   /                                         
  \fi}                                         %
\def\cpto{\mathrel {\vcenter {\baselineskip 0pt \kern 0pt
    \hbox{$CP$} \kern 0pt \hbox{$\longrightarrow$} }}}
\def\cptof{\mathrel {\vcenter {\baselineskip 0pt \kern 0pt
    \hbox{$~CP$} \kern 0pt \hbox{$\longleftrightarrow$} }}}

\begin{document}

\baselineskip=15pt

\preprint{NCTS-PH/1712}

\title{Tests for CPT sum rule and U-spin violation in
Time-dependent CP violation of $B^0_s \to K^+ K^-$ and $B^0 \to \pi^+ \pi^-$}

\author{Xiao-Gang He${}^{1,2,3}$\footnote{hexg@phys.ntu.edu.tw}}
\author{Siao-Fong Li$^{1}$\footnote{r01222004@ntu.edu.tw}}
\author{Bo Ren$^{3}$\footnote{renbo0012@163.com}}
\author{Xing-Bo Yuan$^{2}$\footnote{xbyuan@cts.nthu.edu.tw}}

\affiliation{${}^{1}$CTS, CASTS and Department of Physics, National Taiwan University, Taipei, Taiwan}
\affiliation{${}^{2}$Physics Division, National Center for Theoretical Sciences, Hsinchu, Taiwan}
\affiliation{${}^{3}$INPAC, Department of Physics and Astronomy, Shanghai Jiao Tong University, Shanghai, China}


\begin{abstract}
Recent LHCb data for time-dependent CP violation in $B^0 \to \pi^+\pi^-$ and $B^0_s\to K^+K^-$ show deviations from theoretical predictions. Besides their central values for $\mcC_{K^+K^-}$, $\mcS_{K^+K^-}$ and $\mcA^{\Delta \Gamma}_{K^+K^-}$ violate quantum mechanic CPT invariant sum rule (CPT sum rule) prediction of $\vert \mcC_{K^+K^-}\vert^2 + \vert \mcS_{K^+K^-}\vert ^2 + \vert \mcA^{\Delta \Gamma}_{K^+K^-}\vert ^2 = 1$ (LHCb data imply the sum to be $0.67\pm 0.20$.), their values for $\mcC_{K^+ K^-}= 0.24\pm 0.06\pm {0.02}$ and  $\mcC_{\pi^+ \pi^-} = - 0.24\pm 0.07\pm 0.01$ also show large violation of SU(3) or its U-spin sub-group symmetry (SU(3)/U) relation $\mcC_{K^+ K^-} /\mcC_{\pi^- \pi^+} = -\bar\mcB (B^0 \to \pi^- \pi^+)\tau_{B^0_s}/\bar\mcB (B^0_s \to K^+ K^-)\tau_{B^0}$ (LHCb data imply the ratio of left-side to  right-side to be $4.67\pm 1.88$.) . The LHCb results need to be further confirmed to be taken seriously. We suggest to use time-dependent CP violation in $B_s\to K^0\bar K^0, \pi^+\pi^-, \pi^0\pi^0$ to further test the CPT sum rule. Assuming that the sum rule holds, we propose that violation of the SU(3)/U relation may indicate a large FSI phase difference in the $\pi^+\pi^-$ and $K^+K^-$ re-scattering. We suggest several other U-spin pairs of $B\to PP$ decays  to further test SU(3)/U relations.
\end{abstract}


\maketitle

\section{Introduction}

Recently the LHCb collaboration has measured time-dependent CP violation in $B^0\to \pi^+\pi^-$ and $B^0_s\to K^+K^-$ decays with~\cite{1}
\begin{align}
\mcC_{\pi^+\pi^-} &= -0.24\pm0.07\pm0.01\;,&
\mcS_{\pi^+\pi^-} &= -0.68\pm 0.06\pm0.01\;, 
\nonumber\\
\mcC_{K^+K^-} &= +0.24\pm 0.06\pm0.02\;,&
\mcS_{K^+K^-} &= +0.22\pm0.06\pm0.02\;,
\nonumber\\
\mcA^{\Delta \Gamma}_{K^+K^-}&=-0.75\pm 0.07\pm0.11\;,&&
\end{align}
The above quantities are obtained by measuring time-dependent CP asymmetry $\mathcal{A}(t)$ as a function of time between $B^0_{(s)}$ and $\bar B^0_{(s)}$ meson decaying to a CP eigenstate $f$. The LHCb collaboration determined the above quantities from almost independent features of the decay-time distributions, and hence have almost uncorrelated statistical and systematic uncertainties. Quantum mechanical calculation, assuming CPT invariance, gives~\cite{2}
\begin{eqnarray}
\mcC_f = {1-\vert \lambda_f\vert^2\over 1+\vert \lambda_f \vert^2}\;,\;\;\mcS_f = {2\text{Im}(\lambda_f)\over 1+\vert \lambda_f \vert^2}\;,
\;\;\;\;
{
\mcA^{\Delta \Gamma}_f = - {2\text{Re}(\lambda_f)\over 1+\vert \lambda_f \vert^2}\;,
}
\label{quantities}
\end{eqnarray}
with $\lambda_f = (q/p) (\bar A_f/A_f)$. $A_f$ and $\bar A_f$  are the decay amplitudes for $B^0_{(s)}\to f$ and $\bar B^0_{(s)}\to f$. $q/p$ is a parameter in $B^0_{(s)}-\bar B^0_{(s)}$ mixing whose absolute value is very close to 1.
The quantity $\mcC_f$ is related to direct CP violation $\mcA_\CP(B\to f)=(\vert \bar A_f\vert^2 - \vert A_f\vert^2)/(\vert \bar A_f\vert^2 - \vert A_f\vert^2)$. Since $\vert q/p\vert$ is very close to 1, to a good approximation $\mcA_{\CP}(B \to f) = -\mcC_f$.

Eq.(\ref{quantities}) implies the sum rule
\begin{eqnarray}
\vert \mcC_{f}\vert^2 + \vert \mcS_{f}\vert ^2 + \vert \mcA^{\Delta \Gamma}_{f}\vert ^2 = 1\;. \label{quantum}
\end{eqnarray}
This sum rule holds when the time-dependent CP asymmetry is derived from quantum mechanical evolution of states with CPT invariance. We will refer it as CPT sum rule.

Time-dependent CP asymmetry is not sensitive to $\mcA^{\Delta \Gamma}_f$ in
$B^0 \to \pi^+\pi^-$ because $\Delta \Gamma$ for $B^0 - \bar B^0$ is very small. But, time-dependent CP asymmetry in $B^0_s \to K^+K^-$ can be used to test the CPT sum rule in the above equation.

CP violation in a $B$ decaying into a pair of octet pesudoscalar $PP$ have been extensively studied in the standard model (SM) from model calculations of the values of CP violation to more model independent relations based on symmetry considerations~\cite{3,4,5,gr-l,6,6-a}. One of the interesting one relevant to the LHCb measurement is a relation based on SU(3) or its U-spin sub-group symmetry (SU(3)/U) by exchanging $d$ and $s$ quark in these decays~\cite{5},
\begin{eqnarray}
&&{\mcA_\CP(B^0_s \to K^+ K^- )\over \mcA_\CP(B^0 \to \pi^+ \pi^-)} = - {\overline{\mcB}(B^0\to \pi^+ \pi^-)\tau_{B^0_s}\over \overline{\mcB}(B^0_s \to K^+ K^-)\tau_{B^0}}\;.\label{u-spin1}
\end{eqnarray}
If initial and final state mass differences are included, one should multiply a factor $\lambda_{K^+K^-}^{B^0_s}m_{B^0}/\lambda_{\pi^+\pi^-}^{B^0}m_{B^0_s}$ to correct the phase space difference caused by mass differences, with $\lambda_{ij}^B = [(1-(m_i+m_j)^2/m_B^2)(1-(m_i-m_i)^2/m_B)]^{1/2}$.
Ref.~\cite{fleischer-new} has studied some implications of the above relation to study CKM parameters using the current LHCb data.

In this work, we study implications of the LHCb data for eqs.(\ref{quantum}) and (\ref{u-spin1}). Both equations are violated by the recent LHCb results, with $\vert \mcC_{K^+K^-}\vert^2 + \vert \mcS_{K^+K^-}\vert ^2 + \vert \mcA^{\Delta \Gamma}_{K^+K^-}\vert ^2 = 0.67\pm 0.20$, and the ratio of left-side to right-side of eq.(\ref{u-spin1}) given by $4.67\pm 1.88$, respectively. However, these deviations are about $2\sigma$ significance, and need to be further confirmed. We suggest new measurements to test the relations. We also propose that the violation of SU(3)/U relation may indicate a large Final State Interaction (FSI) phase difference in the $\pi^+\pi^-$ and $K^+K^-$ re-scattering.

\section{Further test of CPT sum rule}

The quantum mechanic nature of the CPT sum rule in eq.(\ref{quantum}) is because that the time-dependent CP asymmetry $\mathcal{A}(t)$ is based on quantum mechanic evolution of the states, $\vert B^0_{(s)}(t)\rangle$, for the $B^0_{(s)} - \bar B^0_{(s)}$ system, with CPT invariance for the mixing matrix and the decay amplitudes~\cite{2}
\begin{eqnarray}
&&\vert B^0_{(s)}(t)\rangle = g_+(t) \vert B^0_{(s)}\rangle -{q\over p} g_-(t) \vert \bar B^0_{(s)}\rangle\;,\nonumber\\
&&\vert \bar B^0_{(s)}(t)\rangle = g_+(t) \vert \bar B^0_{(s)}\rangle -{p\over q} g_-(t) \vert B^0_{(s)}\rangle\;,\nonumber\\
&&g_{\pm}(t) = {1\over 2} (e^{-i m_H t - \Gamma_H t/2} \pm e^{-im_L t -\Gamma_L t/2})\;,
\end{eqnarray}
where $m_{H,L}$ and $\Gamma_{H,L}$ are the masses and lifetimes of the heavier and lighter mass eigenstates.
The above quantum evolution of states leads to the time-dependent CP asymmetry
\begin{eqnarray}
\mathcal{A}(t) = {\Gamma_{\bar B^0_{(s)}\to f }(t) - \Gamma_{B^0_{(s)}\to f}(t)\over  \Gamma_{\bar B^0_{(s)}\to f }(t) + \Gamma_{B^0_{(s)}\to f}(t)} = {-\mcC_f \cos (\Delta m_{(s)} t) + \mcS_f \sin(\Delta m_{(s)}t) \over
\cosh({\Delta \Gamma_{(s)}\over 2}t) + \mcA^{\Delta\Gamma}_f \sinh({\Delta \Gamma_{(s)}\over 2} t)}\;,
\end{eqnarray}
where $\Delta m_{(s)}= (m_H - m_L)_{(s)}$ and {$\Delta \Gamma_{(s)} =(\Gamma_L - \Gamma_H)_{(s)}$} in the $B^0_{(s)} - \bar B^0_{(s)}$ system. The quantities of $\mcC_f$, $\mcS_f$ and $\mcA^{\Delta \Gamma}_f$ are given in eq.(\ref{quantities}). The sum rule in eq.(\ref{quantum}) is therefore also a test of quantum mechanics with CPT symmetry.

The recent LHCb data would lead to
\begin{eqnarray}
\vert \mcC_{K^+K^-}\vert^2 + \vert \mcS_{K^+K^-}\vert ^2 + \vert \mcA^{\Delta \Gamma}_{K^+K^-}\vert ^2 = 0.67\pm 0.20\;.
\end{eqnarray}
The central value seems to violate CPT sum rule~\cite{1,fleischer-new}. If confirmed, it has far reaching implications for quantum field theory. One of course notices that the violation is
less than 2$\sigma$ and one cannot draw a firm conclusion. One has to wait more accurate experimental data to decide.

Quantum mechanic evolutions of states and CPT invariance principles have been tested to great precision in many other systems~\cite{2} and it is difficult to  invent a consistent theory which violates these principles. We will not attempt to build a theoretical model to explain violation of CPT rum rule, but to suggest our experimental colleagues to have more data to verify their results and also to carry out an analysis imposing the CPT sum rule to allow detailed studies of implications using quantum theory with CPT symmetry.

One may wonder whether to $B^0_s - \bar B^0_s$ really form a close system of mixing. If there is a mixing with some other sector (or sectors) with the correct quantum numbers, the sum rule may change. We do not have a good candidate to choose from because the mass of the candidate system should have a mass very close to $B_s^L$ and $B_s^H$. We will not consider modify the sum rule here, but to see if one can test the sum rule using other decays. We find that this is indeed possible by measuring time-dependent CP asymmetries in
\begin{eqnarray}
B^0_s \to K^0 \bar K^0\;,\;\;B^0_s \to \pi^+ \pi^-\;,\;\;B^0_s \to \pi^0 \pi^0\;.
\end{eqnarray}

The branching ratios for $B^0_s \to \pi^+ \pi^-$ and $B^0_s \to \pi^0 \pi^0$ come from annihilation contributions ($A_i$ type of amplitudes) which are expected to be small~\cite{5,6} shown in Table \ref{table1}. However, $B^0_s \to \pi^+ \pi^-$ has already be measured with $\bar\mcB (B^0_s \to \pi^+ \pi^-)= (0.671\pm 0.083)\times 10^{-6}$~\cite{2}. The branching ratio of $B^0_s \to \pi^0 \pi^0$ is predicted to be smaller by a factor of 2 compared with that for  $B^0_s \to \pi^+ \pi^-$ by isospin symmetry which can be seen from Table \ref{table1}. This and the fact that neutral pion is more difficult to measure compared with charged pion make the measurements of relevant quantities for $B_s^0 \to \pi^0\pi^0$ much more difficult. The branching ratio of $B^0_s \to K^0 \bar K^0$ has also been measured to be $(19.6^{+9.7}_{-9.3})\times 10^{-6}$~\cite{2}. Unfortunately CP violation has not been observed in these decays. Theoretical estimations based on QCDF~\cite{8} give small CP violation in these decays with
$\mcC_{K^0\bar K^0}(B_s)\approx -0.40\%$, $\mcS_{K^0 \bar K^0}(B_s) \approx  0.4\%$,
and $\mcC_{\pi^+\pi^-}(B_s) \approx 0$, $\mcS_{\pi^+\pi^-}(B_s) \approx 15 \%$. These estimates show that CP violation are small for these decays  and difficult to measure. One will measure $\vert \mcA^{\Delta \Gamma}\vert^2$ close to 1.
For the pQCD method~\cite{9}, it gives $\mcC_{K^0\bar K^0}(B_s)\approx 0$, $\mcS_{K^0 \bar K^0}(B_s) \approx 4\%$, and $\mcC_{\pi^+\pi^-}(B_s) \approx 1.2\%$, $\mcS_{\pi^+\pi^-}(B_s) \approx 14\% $.

The smallness of CP violation in $\mcC_f$ for
$B_s^0\to K^0\bar K^0$ is due to small tree contribution when using QCDF. This is an accidental cancellation among SU(3) amplitude terms~\cite{6} $C^T_{\bar 3} - C^T_6 - C^T_{\overline{15}} = 0$. This is, however, not generally true if one also include FSI effects. The smallness of CP violation in $B_s^0\to \pi^+\pi^-$ is due to that the decay amplitudes are annihilation type, only $A^i_j$ type of $SU(3)$ amplitudes. This is where the QCDF and pQCD methods become unreliable~\cite{8,wz, 9}. Model independent SU(3) global fitting on the other hand give~\cite{6-a}
$\mcC_{\pi^+\pi^-}(B_s) =\bigl( 16.1_{-1.6}^{+1.9} \bigr)\%$, {$\mcS_{\pi^+\pi^-}(B_s) =\bigl ( -2.3 _{-2.4}^{+2.5} \bigr)\%$}, $\mcA^{\Delta \Gamma}_{\pi^+\pi^-}(B_s)= \bigl( -98.6 _{-0.2}^{+0.4}\bigr)\%$ and $\mcC_{K^0 \bar K^0 }(B_s) = \bigl (-0.9 _{-0.5}^{+0.5} \bigr)\%$, {$\mcS_{K^0\bar K^0}(B_s) =\bigl (-3.5_{-0.5}^{+0.5} \bigr)\%$}, $\mcA^{\Delta \Gamma}_{K^0 \bar K^0}(B_s)=\bigl (-99.9_{-0.0}^{+0.0}\bigr ) \%$\footnote{We have corrected an error for in~\cite{6-a} for $B_s^0 \to K^0 \bar{K^0}$ calculation.}. With more accurate data, the CPT sum rule can be tested with these decay modes. We encourage our experimental colleagues to carry out relevant measurements.

\section{Break down of SU(3)/U relation}

We now study the implications of the LHCb data~\cite{1} on
$\mcC_{\pi^+ \pi^-} = -0.24\pm 0.07\pm 0.01$ and $\mcC_{K^+ K^-}=0.24\pm 0.06\pm {0.02}$. In this study we will assume CPT sum rule in eq.(\ref{quantum}) holds. Imposing this condition, the values obtained by the LHCb for $\mcC_{\pi^+\pi^-}$ and $\mcC_{K^+K^-}$ may change. Here we will take the face values for these quantities to study implications for SU(3)/U violation.

The decay modes $B^0 \to \pi^+\pi^-$ and $B^0_s \to K^+K^-$ are related by SU(3)/U. If this symmetry is a good one, there are some relations among the $U$-spin related pairs and have been used to test the SM and also determine parameters in the model. The decay amplitudes for the two decays in question can be parameterized as
\begin{align}
A(B^0 \to \pi^+\pi^-) &= V_{ub}^*V_{ud} T_d + V_{tb}^*V_{td} P_d\;,\nonumber\\
A(B^0_s \to K^+K^-) &= V_{ub}^*V_{us} T_s + V_{tb}^*V_{ts} P_s\;, \label{amplitudes}
\end{align}
and their corresponding anti-B decay amplitudes are dictated by CPT symmetry to be
\begin{align}
A(\bar B^0 \to \pi^+\pi^-) &= -(V_{ub}V_{ud}^* T_d + V_{tb}V_{td}^* P_d)\;,\nonumber\\
A(\bar B^0_s \to K^+K^-) &= -(V_{ub}V^*_{us} T_s + V_{tb}V_{ts}^* P_s)\;, \label{amplitudes-a}
\end{align}
where $V_{ij}$ are the CKM matrix elements.  $T_i$ and $P_i$ are the tree and penguin amplitudes with CP conserving phases
$\phi_{T_i}$ and $\phi_{P_i}$, respectively. These decay amplitudes in terms of $SU(3)$ invariant amplitudes are given in Table \ref{table1}.
One can also express the amplitudes using diagram approach.
In the SU(3)/U limit, $T_d = T_s$ as can be seen from Table \ref{table1}.

Using the fact that $\text{Im}(V_{ub}^*V_{ud}V_{tb}V_{ud}^*) = - \text{Im}(V_{ub}^*V_{us}V_{tb}V_{us}^*)$,  one obtains
\begin{eqnarray}
&&{\mcA_\CP(B^0_s \to K^+ K^- )\over \mcA_\CP(B^0 \to \pi^+ \pi^-)} = - r_c {\overline\mcB (B^0\to \pi^+ \pi^-)\tau_{B^0_d}\over \overline\mcB (B^0_s \to K^+ K^-)\tau_{B^0}}\;,\nonumber\\
&&r_c = {\text{Im}(T_sP_s^*)\over \text{Im}(T_d P_d^*)} = {\vert T_s\vert \vert P_s\vert \sin(\phi_{T_s} - \phi_{P_s})\over \vert T_d\vert \vert P_d\vert \sin(\phi_{T_d} - \phi_{P_d})}\;.
\end{eqnarray}

\begin{table}
\caption{The tree amplitudes $T_i$ in terms of SU(3) invariant amplitudes for $B\to PP$ where $B$ is one of the $B^+$, $B^0$, $B^0_s$ and $P$ is one of the pions and Kaons. Replacing $T_i$ by $P_i$ one obtains the penguin amplitudes. }
\footnotesize
\begin{eqnarray}
\begin{array}{l}
\left.
\begin{array}{l}
\Delta S = 0\\
T^{B^0_s}_{\pi^+ K^-}(d)=
C^T_{\bar 3}  + C^T_6 -  A^T_{\overline {15}}
+3 C^T_{\overline {15}}\\
T^{B^0}_{\bar K^0 K^0}(d)=
2A^T_{\bar 3} + C^T_{\bar 3} - C^T_6
-3A^T_{\overline {15}} - C^T_{\overline {15}})\\
T^{B^0_s}_{\pi^0 \bar K^0}(d)=-{1\over \sqrt{2}}
( C^T_{\bar 3}  + C^T_6 -  A^T_{\overline {15}} -  5 C^T_{\overline {15}})\\
T^{B^+}_{\bar K^0 K^+}(d)= C^T_{\bar 3}  - C^T_6 +3  A^T_{\overline {15}} - C^T_{\overline {15}}\\
T^{B^0}_{\pi^+ \pi^-}(d)= 2A^T_{\bar 3} + C^T_{\bar 3}  + C^T_6 + A^T_{\overline {15}} + 3C^T_{\overline {15}}\\
T^{B^0}_{K^- K^+}(d)= 2( A^T_{\bar 3} + A^T_{\overline {15}})\\
T^{B^0}_{\pi^0 \pi^0}(d)= {1\over \sqrt{2}}(2A^T_{\bar 3} + C^T_{\bar 3}  + C^T_6 + A^T_{\overline {15}} -5 C^T_{\overline {15}})\\
T^{B^+}_{\pi^0 \pi^+}(d)={8\over \sqrt{2}} C^T_{\overline {15}}\\
\end{array}
\right.
\left.
\begin{array}{l}
\Delta S = -1\\
T^{B^0}_{K^+ \pi^-}(d)= C^T_{\bar 3}  + C^T_6 - A^T_{\overline {15}} + 3C^T_{\overline {15}}\\
T^{B^0_s}_{K^0 \bar K^0}(d)=
2 A^T_{\bar 3} + C^T_{\bar 3}  - C^T_6 - 3 A^T_{\overline {15}}
- C^T_{\overline {15}}\\
T^{B^0}_{K^0 \pi^0 }(d)= - {1\over \sqrt{2}}(C^T_{\bar 3}  + C^T_6 - A^T_{\overline {15}} -5 C^T_{\overline {15}})\\
T^{B^+}_{K^0 \pi^+}(d)=C^T_{\bar 3}  - C^T_6 +3  A^T_{\overline {15}} - C^T_{\overline {15}}\\
T^{B^0_s}_{K^+ K^-}(d)= 2 A^T_{\bar 3} + C^T_{\bar 3}  + C^T_6 +  A^T_{\overline {15}}
+3 C^T_{\overline {15}}\\
T^{B^0_s}_{\pi^+ \pi^-}(d)= {2( A^T_{\bar 3} +  A^T_{\overline {15}})}\\
T^{B^0_s}_{\pi^0 \pi^0}(d)= {\sqrt{2}( A^T_{\bar 3} +  A^T_{\overline {15}})}\\
T^{B^+}_{K^+ \pi^0 }(d)= {1\over \sqrt{2}}(C^T_{\bar 3}  - C^T_6 +3  A^T_{\overline {15}} +7 C^T_{\overline {15}})\\
\end{array}
\right.
\end{array}
\nonumber
\end{eqnarray}
\label{table1}
\end{table}

\begin{table}[!t]
\caption{Experimental results for $\overline{\mcB}(B\to PP)$ and $\mcA_{\CP}(B\rightarrow PP)$ from the HFAG~\cite{7} and PDG~\cite{2}. The sign ``\protect\ec'' indicates that no information is available for the relevant observable.}
\label{table2}
\vspace{1em}
\begin{tabular}{r | c c c | c c c }
&
mode & $\overline{\mcB}\, [10^{-6}]$ & $\mcA_{\rm CP} \,[10^{-2}]$
&
mode & $\overline{\mcB}\, [10^{-6}]$ & $\mcA_{\rm CP} \,[10^{-2}]$
\\\hline

P1)& 
$B^{0}_{s}\rightarrow K^{-}\pi^{+}$ &  $5.5\pm 0.5$ & $26\pm 4$ 
&
$B^{0}\rightarrow K^{+}\pi^{-}$& $19.57^{+0.53}_{-0.52}$ & $-8.2\pm0.6$
\\

P2)&
$B^{0}\rightarrow \overline{K}^{0}K^{0}$ & $1.21\pm 0.16$ & $-0\pm 40$
&
$B^{0}_{s}\rightarrow K^{0}\overline{K}^{0}$ &$19.6^{+9.7}_{-9.3}$ & \ec
\\

P3)&
$B^{0}_{s}\rightarrow \overline{K}^{0}\pi^{0}$ & \ec & \ec
&
$B^{0}\rightarrow K^{0}\pi^{0}$& $9.93\pm 0.49$ & $-0\pm 13$
\\

P4)&
$B^{+}\rightarrow K^{+}\overline{K}^{0}$ &  $1.32\pm 0.14$  & $-8.7\pm 10.0$
&
$B^{+}\rightarrow K^{0}\pi^{+}$ & $23.79\pm 0.75$ & $-1.7\pm 1.6$ 
\\

P5)&
$B^{0}\rightarrow \pi^{+}\pi^{-}$ & $5.10\pm 0.19$ & $24 \pm 7\pm 1$~\cite{1}
&
$B^{0}_{s}\rightarrow K^{+}K^{-}$ & $24.8\pm1.7$ & $-24\pm 6\pm 2$~\cite{1}
\\

&
& & $27 \pm 4$\qquad\;\;\;\;
&
& & \;\;\;
\\

P6)&
$B^{0}\rightarrow K^{+}K^{-}$ & $0.111\pm0.0565$ & \ec
&
$B^{0}_{s}\rightarrow \pi^{+}\pi^{-}$ & $0.671\pm 0.083$ & \ec
\\

&
& & 
&
$B^{0}_{s}\rightarrow\pi^{0}\pi^{0}$& \ec & \ec
\end{tabular}

\end{table}

In the SU(3)/U limit, $r_c=1$ and the above equation reduces to eq.(\ref{u-spin1}). This relation provides a good test of $SU(3)/U$ in the SM. Deviation of $r_c$ from 1 is a measure of SU(3)/U violation.

Using the branching ratios $\overline{\mcB}(B^0\to \pi^+\pi^-) = (5.10\pm 0.19)\times 10^{-6}$, $\overline{\mcB}(B^0_s \to K^+ K^-)=(24.8\pm 1.7)\times 10^{-6}$~\cite{2},
and the new LHCb data on CP violation values for $\mcA_\CP$~\cite{1}, one obtains
\begin{eqnarray}
r_c =  4.67\pm1.88\;.\label{lhcb-central}
\end{eqnarray}
One sees that eq.(\ref{u-spin1}) is badly violated by the central values.

The HFAG~\cite{7} has combined the results of $\mcA_\CP(B^0\to \pi^+\pi^-)$ from different experiments with the average $\mcA_\CP(B^0\to \pi^+\pi^-) = 0.27\pm0.04$. Using also $\mcA_\CP(B^0_s\to K^+ K^-)$ from recent LHCb data, we obtain
\begin{eqnarray}
r_c=4.15\pm 1.30\;. \label{u-global}
\label{central}
\end{eqnarray}
The situation becomes slightly better, but the SU(3)/U predicted relation is still badly violated by the central values.
One should note that the violation is only at $2.42\,\sigma$ and has to wait future improved data to decide how large
the $SU(3)/U$ violation really is.
If the trend of the central values persist, the SU(3)/U is violated.  It is important to also carry out an analysis imposing the CPT sum rule.
Nevertheless, with the present available data it is interesting to investigate how large modifications are needed and also what are the implications for theoretical models trying to calculate $B$ decay amplitudes and CP violation.

There are several theoretical calculations for CP violations in these two decays including certain $SU(3)$ breaking effects~\cite{5,8,9,dhv,Williamson:2006hb}.  A naive factorization calculation gives~\cite{5}
\begin{eqnarray}
r_c \approx   \frac{\lambda_{KK}^{B_s}/m_{B_s}}{\lambda_{\pi\pi}^{B^0}/m_{B^0}}
\left(\frac{(m^2_{B_s}-m^2_K) f_K F^{B_s \to K}(m_K^2)}{(m^2_{B^0}-m^2_{\pi})f_{\pi}F^{B^0 \to \pi}(m^2_{\pi})}\right)^2\;.
\end{eqnarray}
Using $f_{\pi}=132$MeV, $f_{K}=160$MeV, $F_0^{B_s\to K}=0.24$ and $F_0^{B^0\to \pi}=0.25$, one would obtain $r_c = 1.38$.
The naive factorization show a deviation from $r_c=1$, however it is not enough to explain the experiment results in eq.(\ref{u-global}).

QCDF~\cite{8} and  SCET~\cite{Williamson:2006hb} calculations would give for the central value of $r_c$ to be 1.65, 1.02, respectively, still far away from the central value in eq.(\ref{u-global}). pQCD calculations have also been carried out~\cite{9}. Using CP violation in these decays calculated in Ref.\cite{9}, one would obtain a central value for $r_c$ to be 2.74 close to eq.(\ref{u-global}). However, there the theoretical value for the branching ratio of $B_s\to K^+K^-$ is only about half of the experimental one. If using experimental values for the the branching ratios, the value for $r_c$ is again far away from eq.(\ref{u-global}).

\section{Large FSI phases and further tests of SU(3)/U predictions}

We see that within reasonable parameter space, the LHCb value in eq.(\ref{lhcb-central}) and eq.(\ref{central}) cannot be reproduced.
This may be due to that these calculations are mostly dealing with short distance contributions which respect the SU(3)/U to some degree. Long distance interaction contributions may be the source for the large difference which are much harder to calculate. Among the long distance contributions, we suggest that the FSI phase is potentially important for this large SU(3)/U violation effect~\cite{fsi}. This suggestion is inspired by the more precisely measured CP violation which satisfies similar SU(3)/U relation in $B^0 \to K^+ \pi^-$ and $B^0_s \to \pi^+ K^-$.

The decays $B^0 \to K^+ \pi^-$ and $B^0_s \to \pi^+ K^-$ are related in a similar fashion as that for $B^0 \to \pi^+ \pi^-$ and $B^0_s \to K^+ K^-$ by SU(3)/U as can be seen from Table \ref{table1}. In the symmetry limit, one has
\begin{eqnarray}
{\mcA_\CP(B^0 \to K^+ \pi^- )\over \mcA_\CP(B^0_s \to \pi^+ K^-)} = -  r_c {\overline\mcB(B^0_s\to \pi^+ K^-)\tau_{B^0}\over \overline\mcB (B^0 \to K^+ \pi^-)\tau_{B^0_s}}\;,
\end{eqnarray}
with $r_c = 1$.

Using the experimental values listed in table \ref{table2}, $\mcA_\CP(B^0\to K^+\pi^-)=−0.082\pm 0.006$, $\mcA_\CP(B^0_s\to \pi^+K^-)= 0.26\pm 0.04$,
$\mcB(B^0\to K^+\pi^- ) = (19.57^{+0.53}_{-0.52})\times 10^{-6}$, $\tau_{B^0}= 1.520\pm 0.004$ ps,  $\bar\mcB (B^0_s\to \pi^+K^-) = (5.5\pm 0.5)\times 10^{-6}$, $\tau_{B^0_s}= 1.505\pm 0.005$ ps, we obtain
\begin{eqnarray}
r_c = 1.11\pm 0.22\;.
\end{eqnarray}
We see that the SU(3)/U relation is well respected within $1\sigma$ level~\cite{siaofong}.

The SU(3)/U relation holds well for $B^0 \to K^+ \pi^- $ and $B^0_s \to K^- \pi^+$, but not for $B^0\to \pi^+\pi^-$ and $B^0_s \to K^+ K^-$ is puzzling. Why is this so?  To this end,  we notice that in both $B^0 \to K^+ \pi^- $ and $B^0_s \to \pi^+ K^-$, the final states are $K^\pm \pi^\mp$ and are CP conjugate of each other. Their final state phase spaces are the same and also FSI should be similar if the FSI conserves CP which is expected to be so.
But for $B^0 \to \pi^+ \pi^- $ and $B^0_s \to K^+ K^-$ decays, the final state $\pi^+\pi^-$ are very much different than the final state $K^+ K^-$.
The relatively large $s$-quark mass $m_s$ compared with the $u$- and $d$- quark masses $m_{u,d}$ are the source for the SU(3)/U breaking.
In these two decays, the Kaon and pion mass difference cause SU(3)/U breaking causing the short distant decay amplitude differences at about 20\% level.
In addition long distance FSI effects may enhance the SU(3)/U violating effects causing a large difference in inducing FSI phases after getting out short distance effect domain~\cite{fsi}. We propose that deviation of $r_c$ from 1 is mainly caused by different FSI phases of $\phi_{T_i}$ and $\phi_{P_i}$ in $B^0 \to \pi^+ \pi^- $ and $B^0_s \to K^+ K^-$ decays.
In the following, we {adopt a modified Rfit scheme~\cite{Charles:2004jd,Hofer:2010ee,Chang:2014rla}} and fit the $|T|$, $|P|$ and {$\Delta\phi=(\phi_P-\phi_T)$} of $B^0 \to \pi^+ \pi^-$ and $B_s^0 \to K^+ K^-$ decays to the experimental branching ratios and CP violation parameter $\mcC_f$.

\begin{figure}[t]
  \centering
  \includegraphics[width=0.325\textwidth]{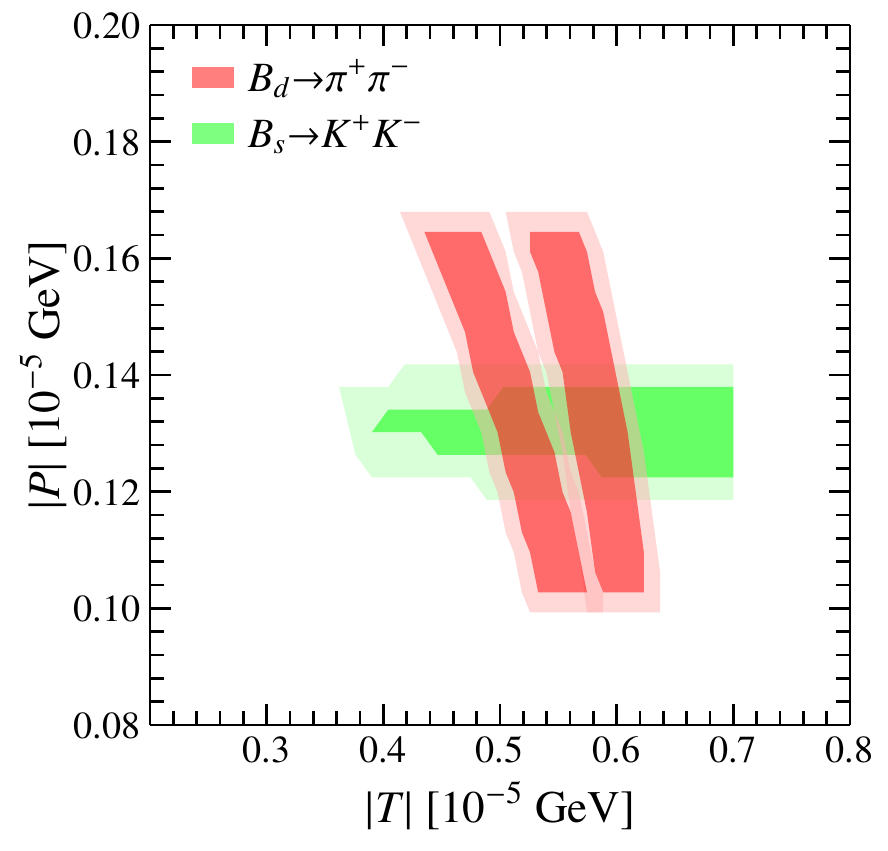}
  \includegraphics[width=0.325\textwidth]{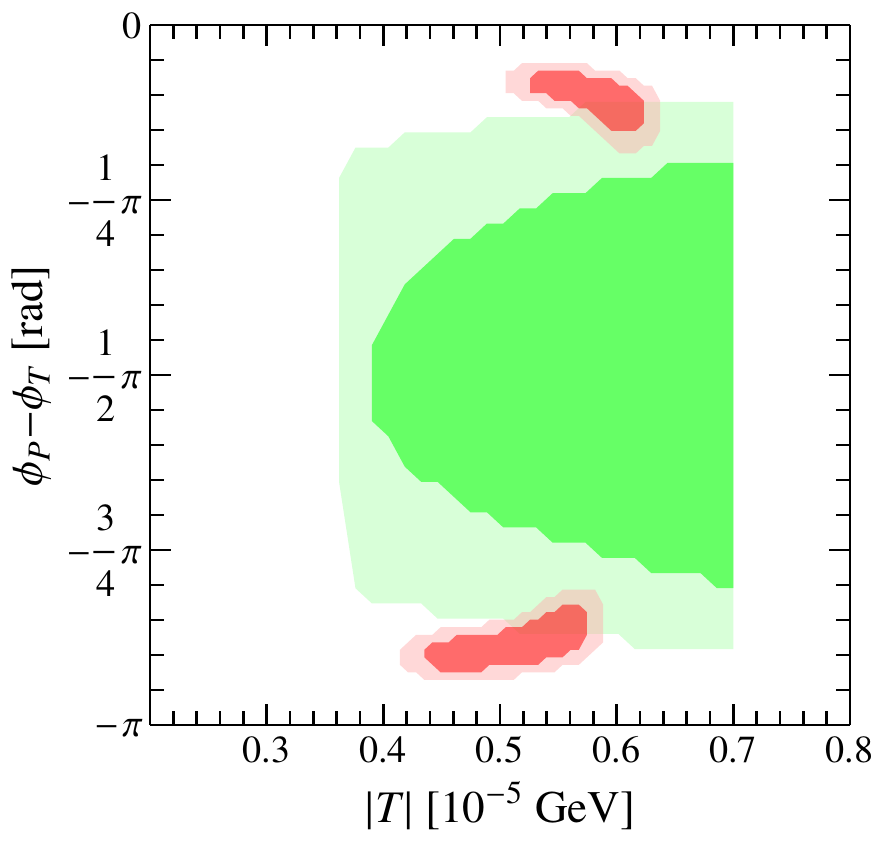}
  \includegraphics[width=0.325\textwidth]{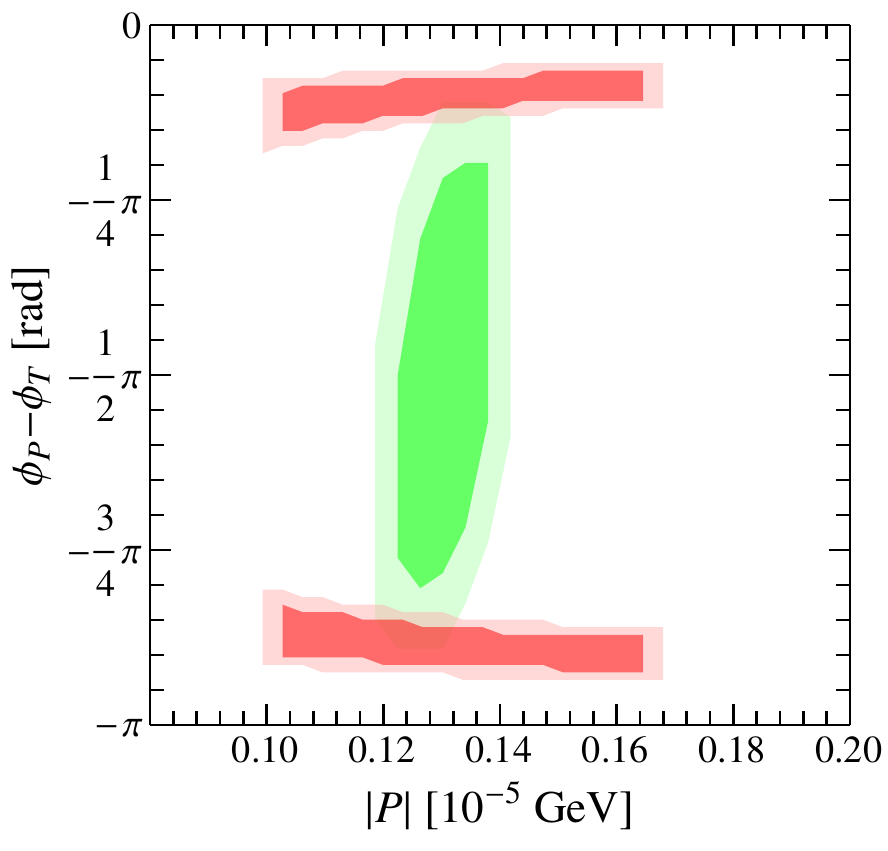}
  \caption{Allowed region of $|T|$, $|P|$ and $(\phi_P-\phi_T)$ for $B^0 \to \pi^+ \pi^-$ (red) and $B_s^0 \to K^+ K^-$ (green) decays under the constraints of branching ratio $\mathcal B$ and CP violation parameter $\mcC_f$. Here, the difference between $|T_d|$ and $|T_s|$ and $|P_d|$ and $|P_s|$ should be less than 20\% is required. The dark and light regions correspond to 68\% CL and 95\% CL, respectively. The best-fit points for $B_d^0 \to \pi^+ \pi^-$ and $B_s^0 \to K^+ K^-$ correspond to $\chi_{\rm min}^2=0.00$ and 0.13, respectively.}
  \label{fig:amp}
\end{figure}

We consider the allowed regions satisfying the difference between $|T_d|$ and $|T_s|$ and $|P_d|$ and $|P_s|$ are less than 20\%, that is $||T_d|-|T_s||/{\rm min}\lbrace |T_d|, |T_s| \rbrace < 20\%$, and similarly for $|P_{s,d}|$. These allowed regions are shown in figure.~\ref{fig:amp}. It can be seen, although $\Delta\phi_s=\Delta\phi_d$ is still allowed at 95\%CL, the current data prefer $\Delta\phi_s\sim -\pi/2$ and $\Delta\phi_d \sim 0$ or $-\pi$, which means an almost imaginary $T_s/P_s$ and real $T_d/P_d$. {In the QCDF approach, the difference between $\Delta \phi_s$ and $\Delta \phi_d$ could origin from different annihilation and spectator scattering contributions, which can be parameterized by the phenomenological parameters $(\rho_A^{i,f},\phi_A^{i,f})$~\cite{Beneke:2001ev,8,Chang:2014rla}. After fixing $\rho_A^{i,f}$ to the solution of the global fit of all $B_{u,d,s}$ decays~\cite{Chang:2014rla}, we find the allowed region of $\rho_A^f$ by $\mcC_{K^+K^-}(B_s)$ is consistent with the one obtained from the global fit only at about $2\,\sigma$ level, which would be a hint of large U-spin violation in these contributions. In addition, the QCDF predictions $\mcC_{K^+K^-}(B_s)=\bigl(11.6_{-0.4-0.4}^{+0.4+0.4}\bigr ) \%$~\cite{Chang:2014rla} and $\bigl( 7.7_{-1.2-5.1}^{+1.6+4.0}\bigr)\%$~\cite{8}, which are based on different $(\rho_A^{i,f},\phi_A^{i,f})$ choices, deviate from the LHCb results at about $2\,\sigma$ level. More theoretical and experimental progresses are needed to clarify this possible puzzle.}

It is well known from $K\to \pi\pi$ decays that large FSI phases exist. There are some analysis on possible large FSI phase in $B \to PP$~\cite{B-isospin} and $D\to PP$~\cite{D-u-violation} decays. In particular it has been shown that in $D^0 \to \pi^+\pi^-$ and $D^0 \to K^+K^-$, the SU(3)/U breaking effects are sizable~\cite{D-u-violation}. Since the penguin contributions to these decays are highly suppressed by GIM mechanism, the SU(3)/U breaking measure can be defined by using their branching ratios and relevant CKM matrices as
\begin{align}
  \tilde r_c(K^+\pi^-, K^- \pi^+)&=\frac{\bar{\mathcal B} (D^0 \to K^+ \pi^-)}{\bar{\mathcal B}(D^0 \to K^- \pi^+)} \cdot \frac{|V_{cs}^*V_{ud}|^2}{|V_{cd}^*V_{us}|^2}, \nonumber\\
  \tilde r_c(K^+K^-, \pi^+\pi^-)&=\frac{\bar{\mathcal B}(D^0 \to K^+ K^-)}{\bar{\mathcal B}(D^0 \to \pi^+ \pi^-)} \cdot \frac{|V_{cd}^*V_{ud}|^2}{|V_{cs}^*V_{us}|^2}.
\end{align}

In the SU(3)/U limit, both $\tilde r_c$ above are equal to 1. Using experimental values for the branching ratios, we obtain $\tilde r_c(K^+\pi^-, K^- \pi^+)\approx 1.3$ and  $\tilde r_c(K^+K^-, \pi^+\pi^-)\approx 2.8$. Similar as the $B$ meson decays, the decays with $K^+ K^-$ and $\pi^+ \pi^-$ final states exhibit larger SU(3)/U violations than that with $K^+\pi^-$ and $K^- \pi^+$ modes.  Our assumption that there is large SU(3)/U violating effect due to FSI in $B^0\to \pi^+\pi^-$ and $B^0_s \to K^+ K^-$  is plausible.

Unfortunately these phases are difficult to calculate. We will not attempt to calculate these FSI phases. Instead, we will concentrate on how the proposal made here can be further tested by more experimental data on CP violation for other possible SU(3)/U related pairs. We list the pairs for $B \to PP$ decays  with which the amplitudes are related by eq.(\ref{amplitudes}) in Table \ref{table2}. These are the pairs named P1), P2), P3), P4), P5) and P6).
From Table \ref{table1}, one sees that $B^0_s\to \pi^0\pi^0$ can also be counted as U-spin related pair with $B^0 \to K^+ K^-$.
Some of the branching ratios and CP asymmetries of the above decay modes have been measured. They are listed in Table \ref{table2}.

In the SU(3)/U limit, $T_d = T_s$ and $P_d = P_s$ and their corresponding $r_c$ are all equal to 1.
Including SU(3)/U breaking effects $r_c$ will be deviate from 1. Depending on the final states of the decay modes, one expects different values for $r_c$.

We have seen from CP violation in $B^0_s \to \pi^+K^-$ and $B^0 \to K^+\pi^-$, $r_c=1.11\pm 0.22$ for P1) decays
is less than 20\% level because the final states in these two decays are CP conjugate states. One therefore also expects that
$r_c$ for P2) and P3) pairs to about 1.11 or so. Measurements for the pairs in P2) and P3) can therefore provide good tests.

The present LHCb data indicate that the pair in P5) have a large SU(3)/U breaking effect
with $r_c = 3.5\pm 1.06$. The pair in P6) and also the pair $B^0_s \to \pi^0 \pi^0$ and $B^0 \to K^+ K^-$ have similar situation as long as final states are concerned. One expects the SU(3)/U breaking effects to be also large with a $r_c$ similar to that for the pair P6). Naively because that these decay modes are annihilation dominated decays, the branching ratios are expected to be small, but the two branching ratios in P6) have been measured. Experimentally, one still need to measure CP violation to test the relations.

For the two pairs in P3) and P4), one of the final state type is two Kaons and another type is one Kaon and one pion. One may expect that SU(3)/U breaking effect due to FSI is between the above two cases. P4) have small branching ratio for $B^+ \to \bar K^0 K^+$ making it more difficult to measure.

\section{Conclusions}

We have studied implications of the recent data for time-dependent CP violation in $B^0 \to \pi^+\pi^-$ and $B^0_s\to K^+K^-$ from the LHCb. The data show deviations from theoretical predictions.

The square sum of the absolute values of $\mcC_{K^+K^-}$, $\mcS_{K^+K^-}$ and $ \mcA^{\Delta \Gamma}_{K^+K^-}$ is given by $\vert \mcC_{K^+K^-}\vert^2 + \vert \mcS_{K^+K^-}\vert ^2 + \vert \mcA^{\Delta \Gamma}_{K^+K^-}\vert ^2 = 0.67\pm 0.20$ which violates quantum mechanical CPT sum rule prediction of 1. We show that experimental data which can be obtained from time-dependent CP violation in $B_s\to K^0\bar K^0, \pi^+\pi^-, \pi^0\pi^0$ can be used to further test the CPT sum rule.

The LHCb values $\mcC_{K^+ K^-}= 0.24\pm 0.06\pm{0.02}$ and  $\mcC_{\pi^+ \pi^-} = - 0.24\pm 0.07\pm 0.01$ imply
$r_c = [\mcC_{K^+ K^-} /\mcC_{\pi^- \pi^+}]/[- \bar\mcB (B^0 \to \pi^- \pi^+)\tau_{B^0_s}/ \bar\mcB(B^0_s \to K^+ K^-)\tau_{B^0}] = 4.67\pm 1.88$. This $r_c$ value
also deviates SU(3)/U  expected value of 1.0.  Assuming that the CPT sum rule holds, we have shown that violation of the SU(3)/U relation may indicate a large FSI phase difference in the $\pi^+\pi^-$ and $K^+K^-$ re-scattering. We suggest to use the pairs of $U$-spin related $B\to PP$ decays listed in Table \ref{table2} to further test the SU(3)/U relations.

Finally, we would like to mention that similar analysis using~\cite{PV} $B \to PV$ and $B\to VV$, where $V$ is one of the vector octet mesons, can test the CPT sum rule and SU(3)/U relations. We encourage our experimental colleagues to carry out relevant measurements.

\begin{acknowledgments}

We thank useful discussions with Chia-Fong Chang and Yu-Kuo Hsiao, and Tim Gershon. This work was supported in part by MOE Academic Excellent Program (Grant No.~105R891505) and MOST of ROC (Grant No.~MOST~104-2112-M-002-015-MY3), and in part by NSFC of PRC (Grant No.~11575111), and also supported by Key Laboratory for Particle Physics, Astrophysics and Cosmology, Ministry of Education, and Shanghai Key Laboratory for Particle Physics and Cosmology (SKLPPC) (Grant No.~11DZ2260700). We thank the authors of ref.~\cite{Chang:2014rla} for providing their QCDF code for $B\to PP$ decays.

\end{acknowledgments}


\begin{thebibliography}{99}

\bibitem{1} LHCb Collaboration, LHCb-CONF-2016-018.

\bibitem{2}
  C.~Patrignani {\it et al.} [Particle Data Group],
  Chin.\ Phys.\ C {\bf 40}, no. 10, 100001 (2016).

 \bibitem{3}
  D.~Zeppenfeld,
  Z.\ Phys.\ C {\bf 8}, 77 (1981);
  M.~J.~Savage and M.~B.~Wise,
  Phys.\ Rev.\ D {\bf 39} (1989) 3346
   [Erratum-ibid.\ D {\bf 40} (1989) 3127];
   L.~-L.~Chau, H.~-Y.~Cheng, W.~K.~Sze, H.~Yao and B.~Tseng,
  Phys.\ Rev.\ D {\bf 43}, 2176 (1991)
  [Erratum-ibid.\ D {\bf 58}, 019902 (1998)];
  M.~Gronau, O.~F.~Hernandez, D.~London and J.~L.~Rosner,
  Phys.\ Rev.\ D {\bf 50}, 4529 (1994).

 \bibitem{4}
  N.~G.~Deshpande and X.~-G.~He,
  Phys.\ Rev.\ Lett.\  {\bf 75}, 1703 (1995).

  \bibitem{5}
  X.~-G.~He,
  Eur.\ Phys.\ J.\ C {\bf 9}, 443 (1999).

  \bibitem{gr-l}
  M.~Gronau and J.~L.~Rosner,
  Phys.\ Lett.\ B {\bf 482}, 71 (2000);
  H.~J.~Lipkin,
  Phys.\ Lett.\ B {\bf 621}, 126 (2005)
  [hep-ph/0503022].

  \bibitem{6}
X.~G.~He, Y.~K.~Hsiao, J.~Q.~Shi, Y.~L.~Wu and Y.~F.~Zhou,
  Phys.\ Rev.\ D {\bf 64}, 034002 (2001);
 H.~K.~Fu, X.~G.~He, Y.~K.~Hsiao and J.~Q.~Shi,
  Chin.\ J.\ Phys.\  {\bf 41}, 601 (2003);
 H.~K.~Fu, X.~G.~He and Y.~K.~Hsiao,
  Phys.\ Rev.\ D {\bf 69}, 074002 (2004).
  X.~G.~He and B.~Mckellar, arxive: hep-ph/0410098;
  C~W~Chiang, M.~Gronau, J.~Rosner and D.~Suprun, Phys. Rev. D{\bf 70}, 034020(2004);
  C.~W.~Chiang, M.~ Gronau, Z.~Luo, J.~Rosner and D.~Suprun, Phys. Rev. D{\bf 69}, 034001(2004);
  C.~W.~Chiang and Y.~F.~Zhou,
  JHEP {\bf 0612}, 027 (2006)
  [hep-ph/0609128].

  \bibitem{6-a}
  Y. ~ K.~Hsiao, C. ~ F.~Chang and X. ~ G.~He,
  Phys.\ Rev.\ D {\bf 93}, no. 11, 114002 (2016)
  doi:10.1103/PhysRevD.93.114002
  [arXiv:1512.09223 [hep-ph]].

 \bibitem{fleischer-new}
 R.~Fleischer, R.~Jaarsma and K.~K.~Vos,
  Phys.\ Rev.\ D {\bf 94}, no. 11, 113014 (2016)
  [arXiv:1608.00901 [hep-ph]].


\bibitem{8}
  H.~Y.~Cheng and C.~K.~Chua,
  Phys.\ Rev.\ D {\bf 80}, 114008 (2009)
  doi:10.1103/PhysRevD.80.114008
  [arXiv:0909.5229 [hep-ph]];
  H.~Y.~Cheng and C.~K.~Chua,
  Phys.\ Rev.\ D {\bf 80}, 114026 (2009)
  doi:10.1103/PhysRevD.80.114026
  [arXiv:0910.5237 [hep-ph]].

\bibitem{9}
  A.~Ali, G.~Kramer, Y.~Li, C.~D.~Lu, Y.~L.~Shen, W.~Wang and Y.~M.~Wang,
  Phys.\ Rev.\ D {\bf 76}, 074018 (2007)
  [hep-ph/0703162 [HEP-PH]].

 \bibitem{wz}
K.~Wang and G.~Zhu,
  arXiv:1304.7438 [hep-ph];
  J.~Liu, R.~Zhou and Z.~J.~Xiao,
  arXiv:0812.2312 [hep-ph].



 \bibitem{dhv}
M.~A.~Dariescu, N.~G.~Deshpande, X.~-G.~He and G.~Valencia,
  Phys.\ Lett.\ B {\bf 557}, 60 (2003);
M.~Beneke,
  eConf C {\bf 0304052}, FO001 (2003)
  [hep-ph/0308040].


\bibitem{Williamson:2006hb}
  A.~R.~Williamson and J.~Zupan,
  Phys.\ Rev.\ D {\bf 74}, 014003 (2006)
  Erratum: [Phys.\ Rev.\ D {\bf 74}, 03901 (2006)]
  doi:10.1103/PhysRevD.74.014003, 10.1103/PhysRevD.74.03901
  [hep-ph/0601214].


\bibitem{fsi}
N.~G.~Deshpande, X.~-G.~He, W.~-S.~Hou and S.~Pakvasa,
  Phys.\ Rev.\ Lett.\  {\bf 82}, 2240 (1999);
  C.~K.~Chua, W.~S.~Hou and K.~C.~Yang,
  Mod.\ Phys.\ Lett.\ A {\bf 18}, 1763 (2003)
  doi:10.1142/S0217732303011551
  [hep-ph/0210002].

\bibitem{7} Y.Amhis et. al., Heavy Flavor Average group, arXiv:1207.1158 and online update at http://www.slac.stanford.edu/xorg/hfag.


\bibitem{siaofong}
X.~G.~He, S.~F.~Li and H.~H.~Lin,
  JHEP {\bf 1308}, 065 (2013)
  [arXiv:1306.2658 [hep-ph]];
  Y.~Grossman, Z.~Ligeti and D.~J.~Robinson,
  JHEP {\bf 1401}, 066 (2014)
  doi:10.1007/JHEP01(2014)066
  [arXiv:1308.4143 [hep-ph]].
  
\bibitem{Hofer:2010ee}
  L.~Hofer, D.~Scherer and L.~Vernazza,
  JHEP {\bf 1102}, 080 (2011)
  doi:10.1007/JHEP02(2011)080
  [arXiv:1011.6319 [hep-ph]].


\bibitem{Charles:2004jd}
  J.~Charles {\it et al.} [CKMfitter Group],
  Eur.\ Phys.\ J.\ C {\bf 41}, no. 1, 1 (2005)
  doi:10.1140/epjc/s2005-02169-1
  [hep-ph/0406184].


\bibitem{Chang:2014rla}
  Q.~Chang, J.~Sun, Y.~Yang and X.~Li,
  Phys.\ Rev.\ D {\bf 90}, no. 5, 054019 (2014)
  doi:10.1103/PhysRevD.90.054019
  [arXiv:1409.1322 [hep-ph]].
  Q.~Chang, J.~Sun, Y.~Yang and X.~Li,
  Phys.\ Lett.\ B {\bf 740}, 56 (2015)
  doi:10.1016/j.physletb.2014.11.027
  [arXiv:1409.2995 [hep-ph]].

\bibitem{Beneke:2001ev}
  M.~Beneke, G.~Buchalla, M.~Neubert and C.~T.~Sachrajda,
  Nucl.\ Phys.\ B {\bf 606}, 245 (2001)
  doi:10.1016/S0550-3213(01)00251-6
  [hep-ph/0104110].
 M.~Beneke and M.~Neubert,
  Nucl.\ Phys.\ B {\bf 675}, 333 (2003)
  doi:10.1016/j.nuclphysb.2003.09.026
  [hep-ph/0308039].

\bibitem{B-isospin}
P.~Guo, X.~G.~He and X.~Q.~Li,
  Int.\ J.\ Mod.\ Phys.\ A {\bf 21}, 57 (2006)
  [hep-ph/0402262].

\bibitem{D-u-violation}
L.~L.~Chau and H.~Y.~Cheng,
  Phys.\ Lett.\ B {\bf 333}, 514 (1994)
  [hep-ph/9404207];
M.~Gronau and J.~L.~Rosner,
  Phys.\ Lett.\ B {\bf 500}, 247 (2001)
  [hep-ph/0010237].

\bibitem{PV}
N.~G.~Deshpande, X.~-G.~He and J.~-Q.~Shi,
  Phys.\ Rev.\ D {\bf 62}, 034018 (2000).


\end{thebibliography}
\end{document}